\documentclass[twocolumn]{aastex62}

\shorttitle{Predicting the Yield of Potential Venus Analogs}
\shortauthors{Ostberg \& Kane}

\usepackage{hyperref}

\begin{document}

\title{Predicting the Yield of Potential Venus Analogs from {\it TESS} and their Potential for Atmospheric Characterization}

\author{Colby Ostberg}
\affiliation{Department of Earth and Planetary Sciences, University of California, Riverside, CA 92521, USA}
\email{costb001@ucr.edu} 
 
\author{Stephen R. Kane}
\affiliation{Department of Earth and Planetary Sciences, University of California, Riverside, CA 92521, USA}


\begin{abstract}

The transit method is biased toward short orbital period planets that are interior to their host star's Habitable Zone (HZ). These planets are particularly interesting from the perspective of exploring runaway greenhouse scenarios and the possibility of potential Venus analogs. Here, we conduct an analysis of predicted {\it TESS} planet yield estimates produced by \citet{Huang}, as well as the {\it TESS} Object of Interest (TOI) list resulting from the observations of sectors 1--13 during Cycle 1 of the {\it TESS} primary mission. In our analysis we consider potential terrestrial planets that lie within their host star's Venus Zone \citep{Kane2014venus}. These requirements are then applied to a predicted planetary yield from the {\it TESS} primary mission \citep{Huang} and the TOI list, which results in an estimated 259 Venus analogs by the end of the {\it TESS} primary mission, and 46 Venus analogs in the TOI list for sectors 1--13. We also calculate the estimated transmission spectroscopy signal-to-noise ratio (S/N) for Venus analogs from the predicted yield and TOI list if they were to be observed by the Near-Infrared Imager and Slitless Spectrograph (NIRISS) on the James Webb Space Telescope ({\it JWST}), as well as update the S/N cutoff values determined by \citet{Kempton}. Our findings show that the best estimated Venus analogs and TOI Venus analogs with $R_{p} < 1.5 \, R_\odot$ have an estimated transmission spectroscopy S/N $> 40$ while planets with radii $2 \, R_\oplus < R_p < 4 \, R_\oplus$ can achieve S/N $> 100$.

\end{abstract}



\section{Introduction}
\label{intro}

The Transiting Exoplanet Survey Satellite ({\it TESS}) is currently observing our nearest and brightest stellar neighbors in search of transiting exoplanets, and will have observed several hundred thousand stars within the F5--M5 spectral type range by the end of its primary mission \citep{Ricker2015}. For a given stellar flux, planets orbiting M dwarfs will have the highest transit probability, making them prime targets for {\it TESS}. Furthermore, their cool temperatures result in compact Habitable Zones (HZ), such as that of the TRAPPIST-1 system \citep{Gillon2017}. Although M dwarfs yield advantages in terms of transit detection probability, their high levels of activity can create difficulties when attempting to observe planetary atmospheres \citep{Kislyakova2019}, and may catalyze atmospheric erosion \citep{LingLoeb2017a,Zendejas2010,Airapetian2019}. The severity of atmospheric erosion that can be caused by M-type stars has yet to be observed, but {\it TESS} will provide many candidates that will allow insight into the conditions in which atmospheres are rapidly desiccated.

The main goal of the {\it TESS} mission is not to aid the analysis of planet distributions or occurrence rates in our galactic neighborhood, but to discover planets amenable to follow up observations \citep{Ricker2015}. Follow up transit and radial velocity (RV) observations will help to constrain the orbital ephemerides of these objects, which is crucial in the planning of future observations \citep{Kane2009}. RV data also provide mass measurements which are needed to constrain the temperature-pressure profiles of a planet's atmosphere, along with its average density. 

Here we are especially interested in the possibility of follow-up observations of planetary atmospheres via transmission spectroscopy using the Near InfraRed Imager and Slitless Spectrograph (NIRISS) instrument equipped to the James Webb Space Telescope ({\it JWST}). NIRISS is expected to be the work-horse for atmospheric characterization \citep{stevenson2016transiting}. The bandpass of the NIRISS instrument has a range of 1--2.5 microns, but can be used in tandem with the Near Infrared Camera (NIRCam) and the Mid-Infrared Instrument (MIRI) to obtain wider spectral coverage. Extensive work has been done in estimating how well we can expect {\it JWST} to perform when used to characterize exoplanet atmospheres \citep{Barstow2015,BatalhaLine,Beichman2014,Belu2011,Clampin2011,Crouzet2017,Deming2009,Greene2016,Howe2017,Louie2018,Molliere2017}. It has been shown that NIRISS alone is expected to be capable of constraining a variety of atmospheric parameters: H$_2$O mixing ratios, the lower limit for atmospheric pressure at the highest cloud altitude within an uncertainty of 1.7 dex, as well as detect the presence of, and in some situations the mixing ratio of CO, CO$_2$, and CH$_4$ \citep{Greene2016}. These measurements are all possible with one transit, and uncertainties will decrease with each observed transit. However, similar to what has been observed with observations of atmospheres using the Hubble Space Telescope (HST), these measurements become much more arduous when cloud cover is included \citep{Croll2011,Kreidberg2016}. This causes the uncertainties for measurements to rise, but should not impede the ability to constrain molecular mixing ratios, carbon-to-oxygen ratio, [Fe/H], and temperature--pressure (T-P) profiles, especially when utilizing the full spectral coverage of {\it JWST} \citep{Greene2016}.

Several estimates of the planets {\it TESS} will discover ({\it TESS} yields) have been published. One of the more widely accepted predicted yields has been produced by \citet{Sullivan2015} (the "Sullivan yield"), which uses an artificial group of stars in tandem with planetary occurrence rates derived from Kepler \citep{DressingCharb,Fressin2013}. However, discrepancies in the stellar and planetary occurrence rates implemented by \citet{Sullivan2015} have been discovered \citep{Ballard2019,Barclay2018,Bouma2017,Huang}, which affect the accuracy of the yield. In this work we adopt the yield produced by \citet{Huang} (the "Huang yield"), as it utilizes stars from the {\it TESS} Input Catalog (TIC) \citep{Stassun2018} for its stellar population, as well as updated information on {\it TESS} predicted systematic noise levels and multi-planet system occurrence rates. 

Due to the intrinsic sensitivity of transit detection, it is anticipated that the planets discovered by {\it TESS} will yield orbits that place them either within or interior to their respective star’s HZ. The planets interior to their HZ have the potential to lie within the confines of the Venus zone (VZ) \citep{Kane2014venus}. The VZ is the region between the Runaway Greenhouse boundary defined by \citet{Kopparapu2013}, and the distance from a star where the planet would receive 25 times the stellar flux received by the Earth. Characterization of planets within the VZ will lead to clarity of the habitability dichotomy we observe between Earth and Venus \citep{Kane2019}. Inferences of these planets' climates can be made by applying observed atmospheric abundances from {\it JWST} observations into 3-D general circulation models (GCM), similar to work by \citet{Way2016}. Better understanding of climates that can exist within the VZ will help constrain the Runaway Greenhouse boundary, and parameters which caused the divergence in habitability between Earth and Venus. 
In this work, we provide the results of an extensive analysis of the \citet{Huang} predicted {\it TESS} yield. In Section~\ref{methods} we explain our selection of the Huang yield, compare its results to that of the Sullivan yield,  and define the Transmission Spectroscopy Metric. In Section~\ref{results} we calculate the VZ for all stars within the Huang yield and predict the total number of  Venus analogs that {\it TESS} will find. In Section~\ref{discussion} we apply the TSM to our adopted yield to obtain S/N values we can expect to see when attempting to characterize the planetary atmospheres of the Huang yield with the {\it JWST} NIRISS instrument, as was done by \citet{Kempton} with the Sullivan yield. This is used to create updated S/N cutoff values to be used to prioritize which discovered planets are the best candidates for transmission spectroscopy follow up. We provide concluding remarks and prospects for future work in Section~\ref{conclusions}.


\section{Methods}
\label{methods}

\subsection{\textit{TESS} Yield Selection}
\label{yield selection}

There are several predicted {\it TESS} yields that have been published, each with their own unique additions and modifications \citep{Barclay2018,Bouma2017,Huang,Sullivan2015}. \citet{Sullivan2015} has set the precedent for estimating planetary yields, and has been widely accepted  and used by the {\it TESS} community. However, errors have been uncovered in this yield which affect the planet population it produces. More recent yields have since been produced that account for these errors \citep{Barclay2018,Bouma2017,Huang}, of which all have slight differences. We adopt the yield of \citet{Huang} for this work as it contains a number of updates: a refined photometric noise model which lowers the predicted systematic noise floor of {\it TESS} to 40~ppm, accurate stellar parameters acquired using GAIA DR2 data \citep{Andrae2018} and the TIC, and refined multi-planet system estimates \citep{Ballard2016,Zhu2018}. A more in depth analysis of these updates and additional adjustments that were made can be found in the literature. 

\begin{figure*}
  \centering
  \includegraphics[width=\textwidth]{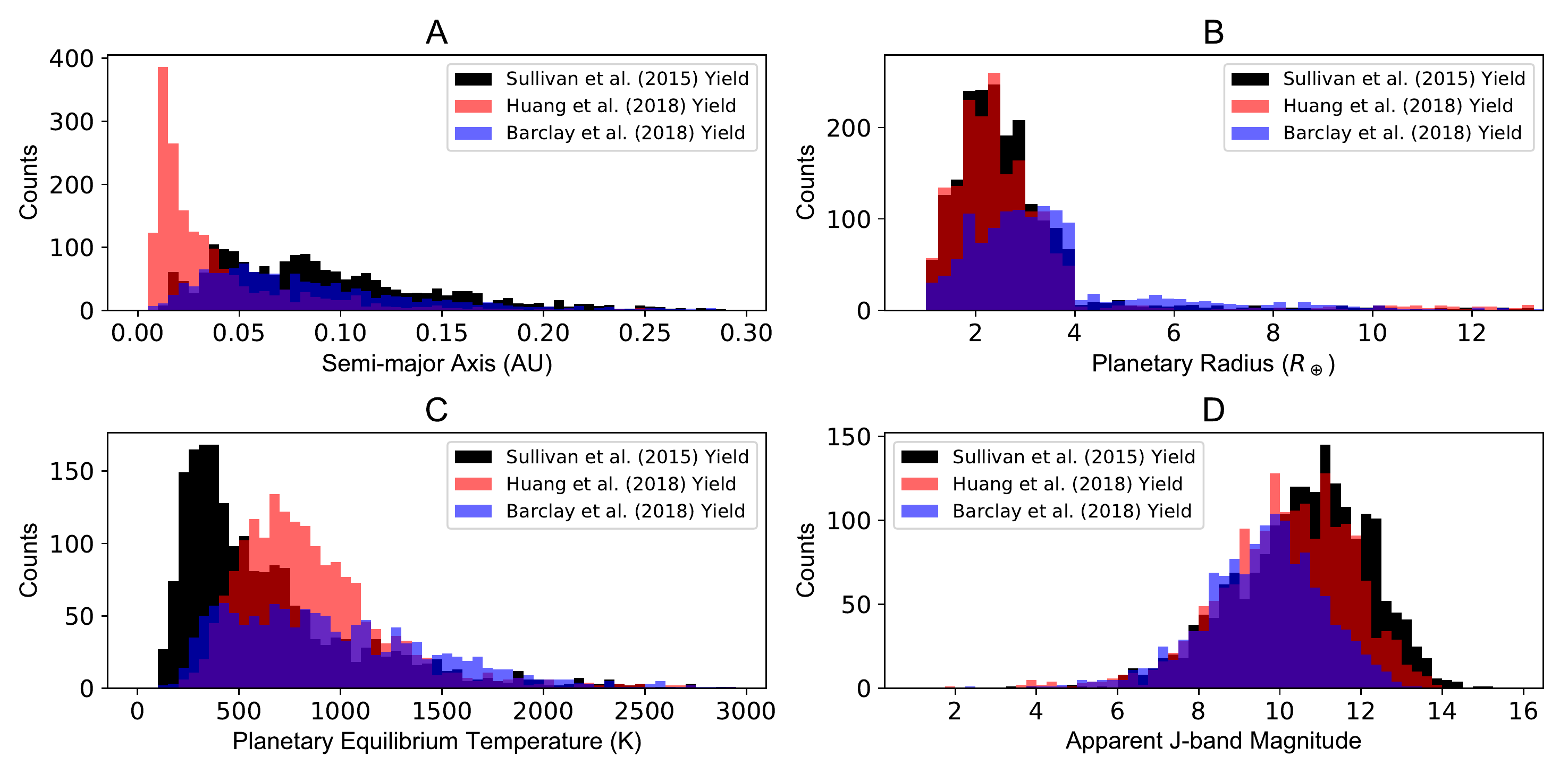}
  \caption{A comparison of the \citet{Sullivan2015}, \citet{Huang}, and \citet{Barclay2018} yields. {\bf A:} The semi-major axis distributions of the three yields. All planets are considered to have circular orbits. {\bf B:} The range of planetary radii produced by each yield. {\bf C:} The differences in planetary equilibrium temperature assuming zero albedo and full heat redistribution. {\bf D:} Apparent J-band magnitudes for all yields.}
  \label{Yieldscomparison}
\end{figure*}

To illustrate how the updated information creates differences in the yield produced, we have compared 4 significant stellar and planetary parameters of the Huang, Sullivan, and Barclay yields (Figure \ref{Yieldscomparison}). The Huang yield predicts the majority of planets from the {\it TESS} mission to have semi-major axes within 0.05 AU, whereas the Sullivan and Barclay yields predicts a more uniform distribution of semi major axes which stretches out to 0.3 AU (Figure~\ref{Yieldscomparison}A).  All yields show similarities in their estimates on the distribution of planetary radii (Figure \ref{Yieldscomparison}B), while slight discrepancy in predicted J-band magnitudes can be seen as the Barclay yield is centered on brighter stars (Figure \ref{Yieldscomparison}D). Resulting from the discrepancy in semi-major axes, the Huang yield boasts a higher density of planets with high equilibrium temperatures (Figure \ref{Yieldscomparison}C). As will be discussed in coming sections, these differences in orbital radii and planetary equilibrium temperature will lead to much larger estimated S/N values for the Huang yield when compared to S/N values calculated by \citet{Kempton} for the Sullivan yield. It should be noted that there is a large disparity between the Barclay yield, and Huang and Sullivan yields when it comes to the total amount of planets expected to be found. The Huang and Sullivan yields predict that target pixels in the TESS primary mission will find a total of 1799 and 1984 planets, respectively. While the Barclay yield predicts a total of 1293 planets. This discrepancy in total planets discovered affects the differences in distributions shown in Figure~\ref{Yieldscomparison}, as well as the difference in Venus analogs predicted, as explained in a following section.


\subsection{Transmission Spectroscopy Metric}
\label{Methods_TSM}

Models of the prospected performance of {\it JWST} have been applied to estimates of {\it TESS} yield to estimate the S/N that can be achieved through transmission spectroscopy \citep{Louie2018}. But the time that is needed to run these models can be substantial, and is impractical when attempting to be timely in comparing the S/N values of a large set of planets. To expedite this process, \citet{Kempton} developed a transmission spectroscopy metric (TSM) which allows one to produce values proportional to the S/N of the spectral features observed during a 10 hour observing run using the NIRISS on JWST. \citep{Louie2018}:
 \begin{equation}
     TSM = (\mathrm{Scale\,Factor}) \times \frac{R_p^3 \, T_{\mathrm{eq}}}{M_p\,R_\star^2} \times 10^{-m_J/5}
 \label{tsm equation}
 \end{equation}
where $T_\mathrm{eq}$ is the planetary equilibrium temperature, $m_J$ is the host star's apparent J-band magnitude; $R_p$ and $M_p$ are the radius and mass of the planet in units of Earth radii and Earth masses, respectively; and $R_\star$ is the radius of the star in solar radii. The scale factor is needed to allow the calculated TSM value to have a 1:1 ratio with the simulated NIRISS S/N produced by \citet{Louie2018}, and differs based on the radius of the planet.

For the sake of simplicity, the mean molecular weights of the planets' atmospheres are assumed based on the planetary radius. Whereas for planets with radius $R_p < 1.5 R_\oplus$, a mean molecular weight of $\mu = 18$ (in units of proton mass), giving these planets have a steam dominant atmosphere. For planets with $R_p > 1.5 R_\oplus$, a hydrogen dominated atmosphere is assumed by using a mean molecular weight of $\mu = 2.3$. Other assumptions that have been made are that all atmospheres are cloud free, and that the masses of the planets can be accurately determined from an empirical mass-radius relationship \citep{ChenKipping}. 


\subsection{Stellar Relationships}
\label{relations}

\begin{figure*}
  \includegraphics[width=\textwidth]{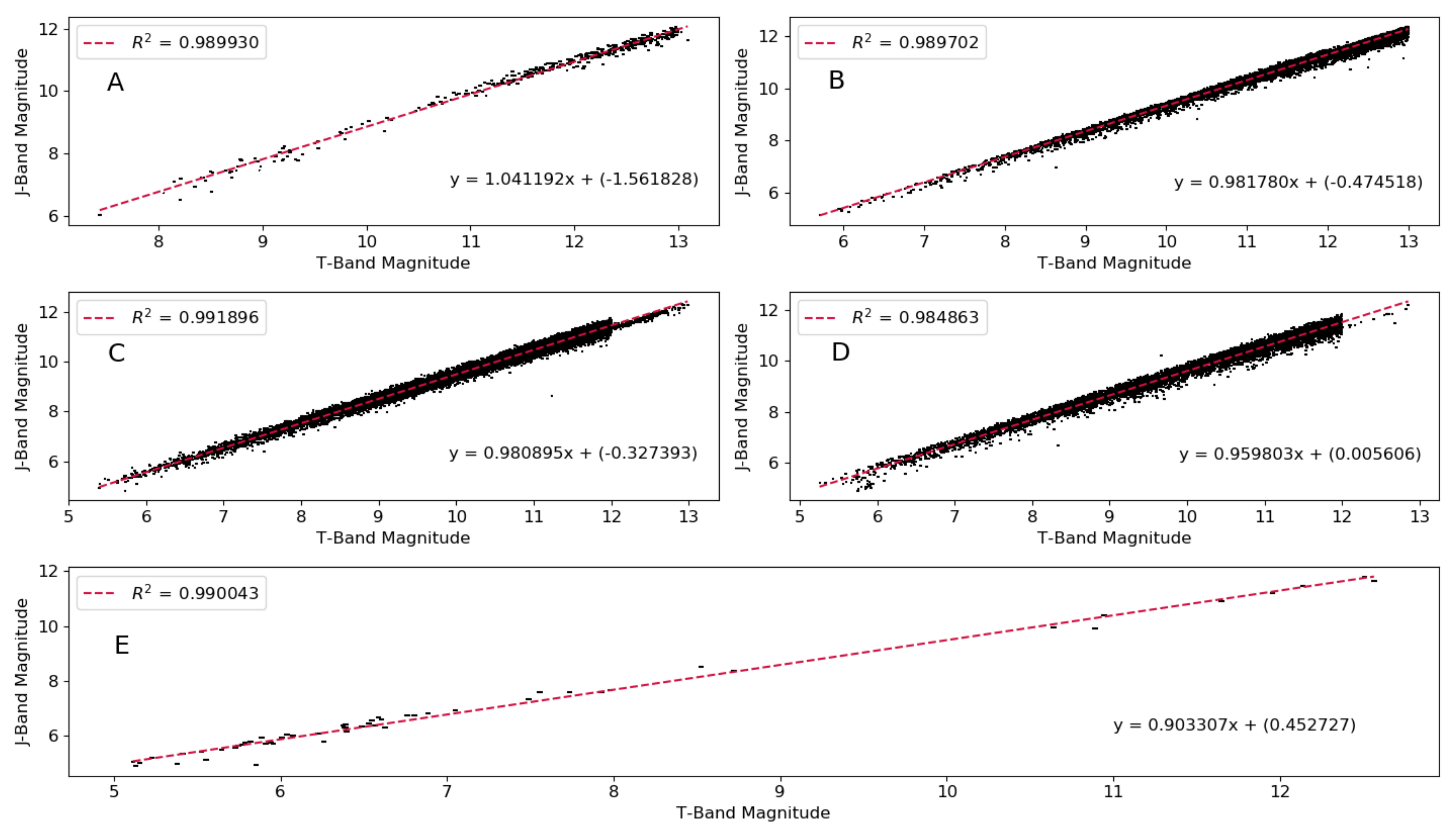}
  \caption{The apparent T- and J-band magnitude relationship for each luminosity bin (A = $0.013~L_\odot$ to $0.13~L_\odot$, B = $0.13~L_\odot$ to $1.30~L_\odot$, C = $1.30~L_\odot$ to $13.0~L_\odot$, D = $13.0~L_\odot$ to $130~L_\odot$, and E = $130~L_\odot$ to $130000~L_\odot$). Each fit's associated goodness of fit value and linear equation with best fit parameters are shown. Error bars associated with the data are included.}
  \label{Mag fit}
\end{figure*}

A problem was encountered when trying to apply the TSM equation to the Huang yield, as the data produced from their simulation included only the apparent {\it TESS}-band (T-band) magnitude of the host star, whereas the apparent J-band magnitude is needed for Equation~\ref{tsm equation}. To surpass this, we gathered both J- and T-band magnitudes for $\sim$45,500 stars on the {\it TESS} Candidate Target List (CTL) available on the MAST website (\url{https://mast.stsci.edu}), and developed a linear relationship between the two. This specific amount of stars resulted from requiring that the uncertainty in  T- and J-band magnitude be less than 0.018, in order to increase the precision of our derived relationship. To better the quality of the fit, it was necessary to create separate relationships based on five luminosity ranges, which in combination cover the entire range of stellar luminosities observed in the Huang yield (Figure~\ref{Mag fit}). Using these relationships we converted the provided T-band magnitudes to the required J-band and were able to determine the TSM values for the entire yield.

\begin{figure*}
  \includegraphics[width=\textwidth]{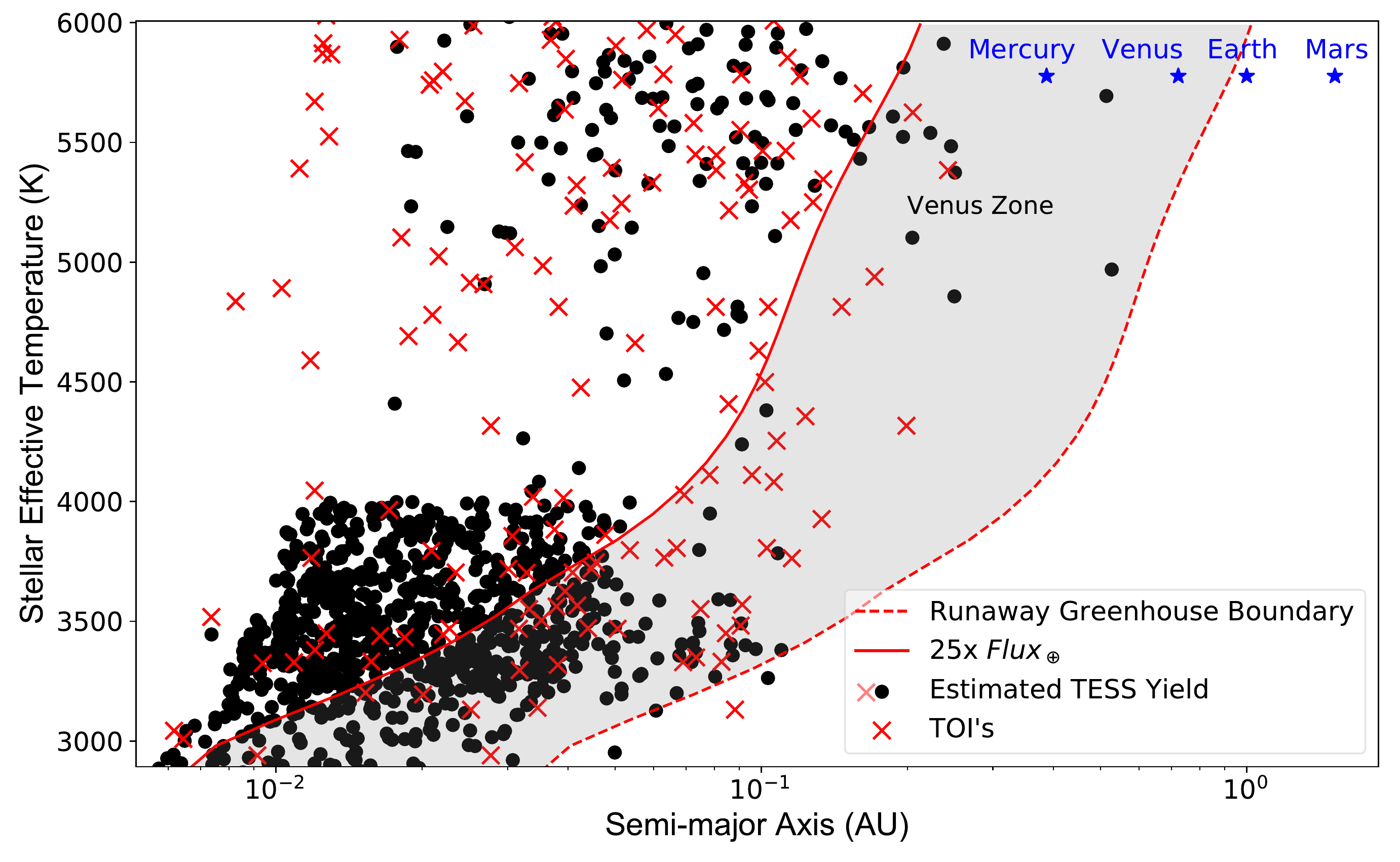}
  \caption{\citet{Huang} planets and TOI planets with radii less than 2.5~$R_\oplus$ with respect to the VZ (grey). The solar system terrestrial planets are also shown as a reference.}
  \label{VZ1}
\end{figure*}

This study also requires the calculation of the Runaway Greenhouse boundary with respect to stellar temperature and orbital semi-major axis. The equation for the HZ boundary as defined by \citet{Kopparapu2013} is expressed as follows:
\begin{equation}
    d = \left( \frac{L_\star/L_\odot}{S_\mathrm{eff}} \right)^{0.5} \ \mathrm{AU}
    \label{dist2hz}
\end{equation}
where $L_\star$ is the luminosity of the host star, $L_{\odot}$ is the luminosity of the sun, and $S_\mathrm{eff}$ is incident stellar flux which has specified values for each HZ boundary. Performing this calculation thus requires an estimate of the stellar luminosity based upon the $T_\mathrm{eff}$ values provided. There are numerous relationships between main sequence stellar parameters \citep[e.g.,][]{Boyajian2012,Eker2015} but relatively few empirically derived relationships between $L_\star$ and $T_\mathrm{eff}$ that may be applied across a broad range of stars. To resolve this we developed a relationship between stellar mass ($M_\star$) and effective temperature using MESA stellar isochrones available on the MIST website (\url{http://waps.cfa.harvard.edu/MIST/}). We fit a 5th degree polynomial of the form $f(x) =  ax^5 + bx^4 + cx^3 + dx^2  + ex + f$, with coefficients $a = -3.16669290 \times 10^{-17}$, $b = 7.35506985 \times 10^{-13}$, $c = -6.70248885 \times 10^{-9}$,  $d = 2.98676177 \times 10^{-5}$, $e = -6.46102582 \times 10^{-2}$, and $f = 54.2221352$. There are range of $T_\mathrm{eff}$ values that are associated with a single main sequence stellar mass, spanning a range of up to 800~K depending on age and metallicity. We assumed a stellar age of $\sim$4~Gyr, based on 4~Gyr being a typical age for field stars where ages have been determined via asteroseismic and rotation period techniques \citep{vansaders2016}. We also assumed an average extinction of zero for all stars, since primary {\it TESS} targets are relatively close, and [Fe/H] values in the range -0.1142--0.0142~dex, to cover a relatively large range of metallicities while avoiding large variations in effective temperature. The polynomial fit to these data described above allowed us to use stellar effective temperature to estimate mass, from which mass-luminosity relationships can be used to determine the luminosity. By combining the derived polynomial with Equation~\ref{dist2hz}, the distance to the HZ boundaries is then solely dependent on $T_\mathrm{eff}$, allowing us to plot the outer boundary of the VZ shown in Figure~\ref{VZ1}. Note that this derived relationship is only used to determine the VZ boundaries and not for extracting stellar parameters for the stars included in this study.

\subsection{Defining Venus Analogs and the Venus Zone}
\label{Venus analog defined}

Much work has been done in estimating the radius value where we would expect planets to transition from terrestrial to gaseous \citep{ChenKipping,Lopez2014,Rogers2015}, all of which agree that the upper limit of the terrestrial regime can be found around $\sim 1.75 \, R_\oplus$. Considering that we do not want to exclude any terrestrial planets in our estimates, we require all planets to be considered Venus analogs to have a radius $R_p < 2.5 \, R_\oplus$. This large upper bound in radius serves as a buffer which accounts for uncertainty of measurements. This is especially important for planets orbiting dimmer stars as the small amount of flux we receive from them will result in large uncertainties in the stellar radii, which directly translates to uncertainties in planetary radii. This buffer will result in the inclusion of some sub-Neptune planets in our Venus analog yield, but will assure that no terrestrial planets are excluded.

The second requirement in our definition of a Venus analog is that the planet has sufficient insolation flux to place it within the boundaries of the VZ \citep{Kane2014venus}. The inner boundary of the VZ is located where a planet would receive flux from its host star equal to 25$\times$ the flux received by Earth. This specific amount of flux is used since it is the same amount of flux that would place Venus on the “cosmic shoreline”, which is the tipping point of where Venus would start to experience severe atmospheric loss \citep{ZahnleCatling}. The outer boundary is the Runaway Greenhouse boundary defined by \citet{Kopparapu2013}, where the flux received would cause surface water on an Earth-like planet to be completely evaporated. This increase in H$_2$O in the atmosphere decreases the amount of outgoing infrared radiation, which triggers severe climate warming. Considering the definitions of the two boundaries, we expect planets in the VZ to lack surface water but still have a considerable atmosphere. It should be noted that these boundaries only take into consideration the effect of incident stellar flux on the planet, however there are many other effects to consider when inferring a planet's climate (e.g. tidal heating, rotation rate, magnetic field, etc.). The purpose of defining Venus analogs in this work is not an attempt to define which planets are completely analogous to Venus, but instead as a target selection tool for planets who would be prime candidates for follow up observations that would allow us to test the hypothesis of the VZ.


\section{Results}
\label{results}

\subsection{Expected Yield of Exo-Venuses}
\label{venus yield}

In applying our conditions for what we define to be a Venus analog, we find that the Huang yield predicts that {\it TESS} will discover 259 planets with radius $R_p < 2.5 \, R_\oplus$, and sufficient stellar flux to be placed in the VZ (Figure~\ref{VZ1}). This number derives only from the predicted planetary yield from {\it  TESS} 2-minute cadence using target pixels, making this a lower bound estimate as it does not account for the planets discovered from full-frame images (FFI) or the extended missions currently being planned. For further predictions on the outputs of the {\it TESS} extended missions, we refer the reader to \citet{Bouma2017} and \citet{Huang}. We also applied our definition of a Venus analog to the {\it TESS} Object of Interest (TOI) list containing data from the observations of sectors 1--13, which produced 46 Venus analogs (Figure~\ref{VZ1}). It is important to note that the TOI's have yet to be confirmed as actual planets, and until follow up observations can verify their presence, our estimate of Venus analogs in the TOI list is tentative. 

\begin{figure*}
  \includegraphics[width=\textwidth,height=10cm]{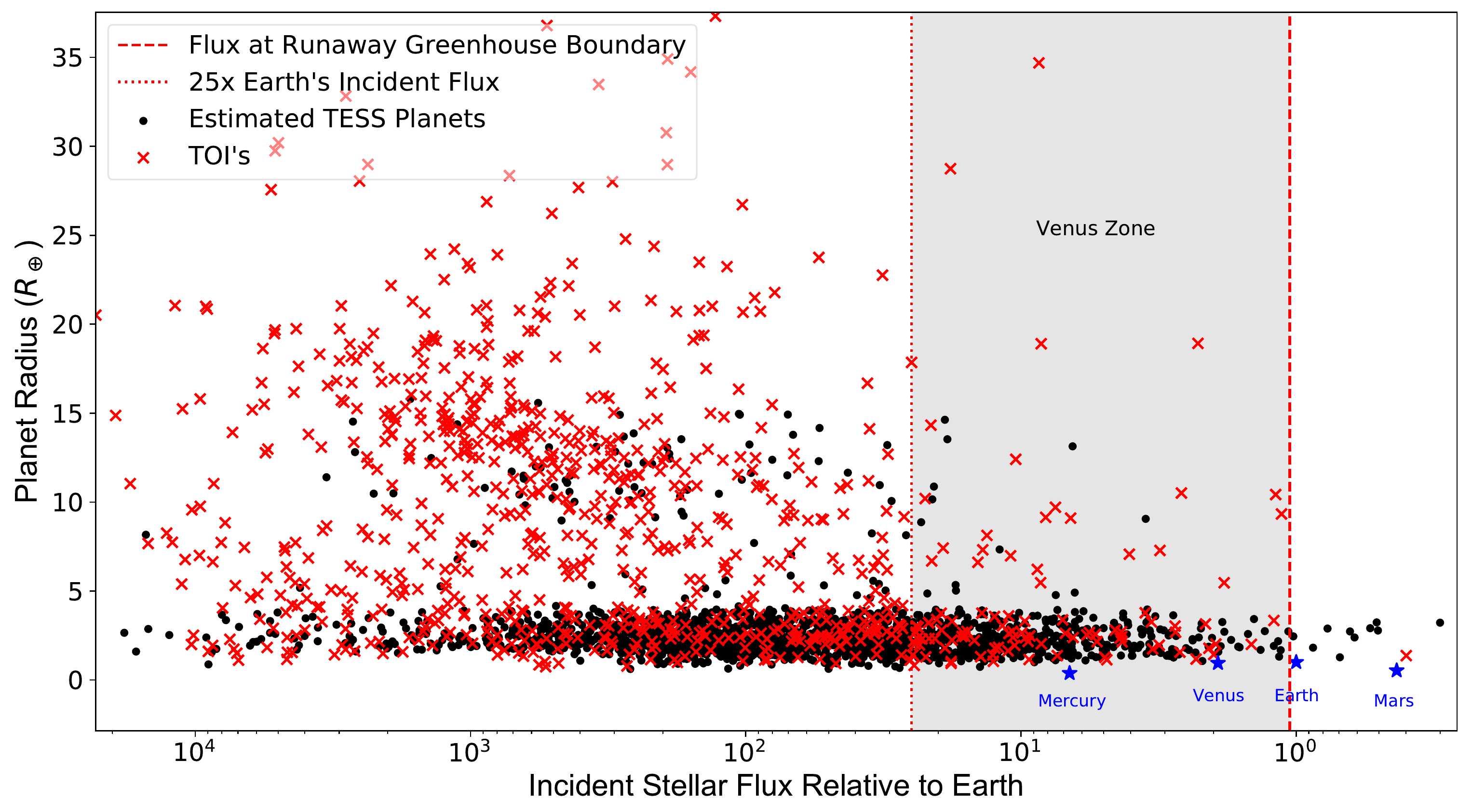}
  \caption{All planets from the \citet{Huang} yield (black dots) and TOI list (red crosses) with respect to the VZ defined by insolation flux. The solar system terrestrial planets are shown for reference.}
  \label{VZ2}
\end{figure*}

To get a better idea on the full range of planet demographics we expect to find in relation to the VZ, Figure~\ref{VZ2} includes planets from the TOI list and Huang yield of all radii in respect to the VZ. It can be seen that majority of the planets predicted by the Huang yield lie interior to the VZ, many of which are in the super-Earth and sub-Neptune range, creating opportunities to statistically fortify the theoretical boundary between terrestrial and gaseous planets \citep{ChenKipping,Lopez2014,Rogers2015}. There is also a several planets that lie near the VZ boundaries, which could provide an opportunity to test the hypotheses from which these boundaries were conceived.


\subsection{TSM Values}
\label{tsm values}

\begin{figure}
  \includegraphics[width=\columnwidth]{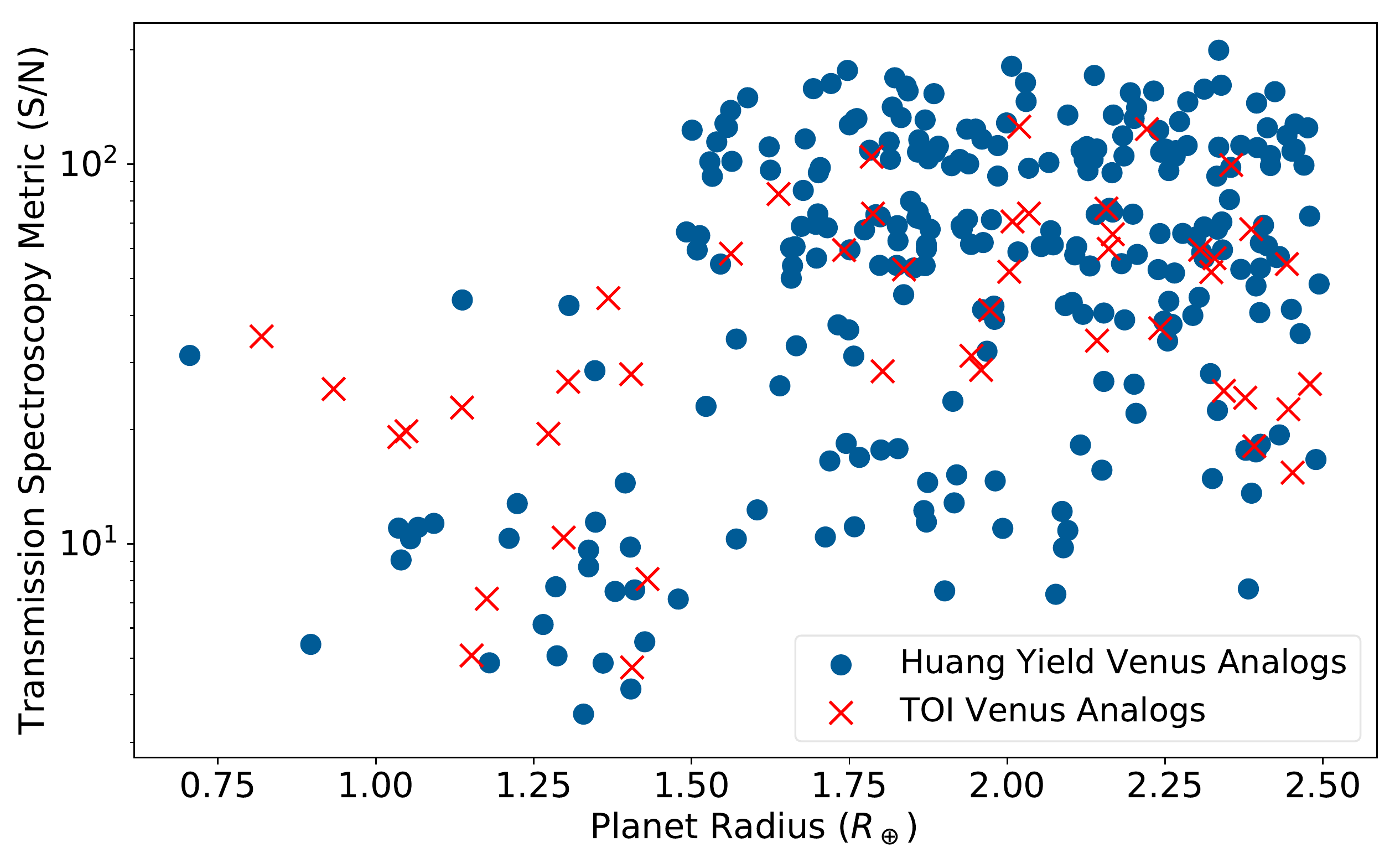}
  \caption{The distribution of S/N values obtained from applying Equation \ref{tsm equation} to Venus analogs in the Huang et al. (2018) yield and TOI list.}
  \label{TSM}
\end{figure}

The purpose of determining the TSM values for planets is to obtain an estimate as to how well we would be able to resolve a planet's atmospheric composition using the NIRISS instrument on {\it JWST} \citep{Kempton}. Figure~\ref{TSM} displays the distribution of TSM values we can expect to find for Venus analogs from the TOI list and Huang yield. The decrease in TSM values for planetary radii $< 1.5 \, R_\oplus$ is due to the assumptions regarding the atmospheric compositions, outlined below. One can see that there are several planets from the Huang yield and TOI list with estimated S/N~$> 20$, while it's also observed that there is a significant number of planets with S/N~$> 60$. However many of planets with S/N~$> 60$ have radii which place them near or beyond the theoretical terrestrial boundary. It is important to note that these values would only be relevant for planetary atmospheres that have significant absorption of photons in the near infrared, since they reflect the expected performance of the NIRISS instrument whose bandpass is 1--2.5 microns. The high expected S/N for observations of these planet's atmospheres increases the likelihood of observing the absorption features of prominent molecules. However, a significant caveat is that these values are for clear atmospheres with minimal clouds. If the planets we hope to observe are true Venus analogs, then opacity due to clouds will hinder our ability to peer deep into their atmospheres, and the observed S/N will not reflect the estimated S/N. Instead we would only be able probe the uppermost layers of the atmospheres, resulting in the need to extrapolate the constituents and abundances of the rest of the atmosphere. 

With the inevitable high level of competition for {\it JWST} observing time once it is launched, it is necessary to create a S/N cutoff value to designate the best candidates for atmospheric observations. As postulated by \citet{Kempton}, 300 candidates from the {\it TESS} mission with radii less than $10 \, R_\oplus$ would create a diverse sample to provide evidence for unconfirmed theories in planetary science. To obtain this sample we chose 70, 100, 100, and 30 planets with the highest TSM values for the  $R_p < 1.5 \, R_\oplus$, $1.5 \, R_\oplus < R_p < 2.75 \, R_\oplus$, $2.75 \, R_\oplus < R_p < 4 \, R_\oplus$, and $4 \, R_\oplus < R_p < 10 \, R_\oplus$ radius bins, respectively. This makes the cutoff value for each radius bin the lowest TSM value in that bin (Table~\ref{Table 1}). These values are solely an update to the TSM cutoff values derived from the Sullivan yield, previously done by \citet{Kempton}. 

\begin{table*}[]
    \centering
    \begin{center}
    \begin{tabular}{|c|c|c|c|c|} 
    \hline
    Radius Bin & $R_p < 1.5\, R_\oplus$ &  $1.5 \, R_\oplus < R_p < 2.75 \, R_\oplus$ & $2.75 \, R_\oplus < R_p < 4 \, R_\oplus$ & $4 \, R_\oplus < R_p < 10 \, R_\oplus$ \\ [0.5ex] 
    \hline
    S/N Cutoff & 18 & 176 & 155 & 71 \\
    \hline
    Planets Above Cutoff & 70 & 100 & 100 & 30 \\
    \hline 
    Total Planets in Bin & 217 & 988 & 490 & 41 \\
    \hline
    \end{tabular}
    \end{center}
    \caption{Cutoff values for each radius bin which would create a statistical sample of 300 planets from {\it TESS} target pixel discoveries in its primary mission}
    \label{Table 1}
\end{table*}
 

\subsection{Yield Degeneracy}

\begin{figure*}
  \includegraphics[width=\textwidth]{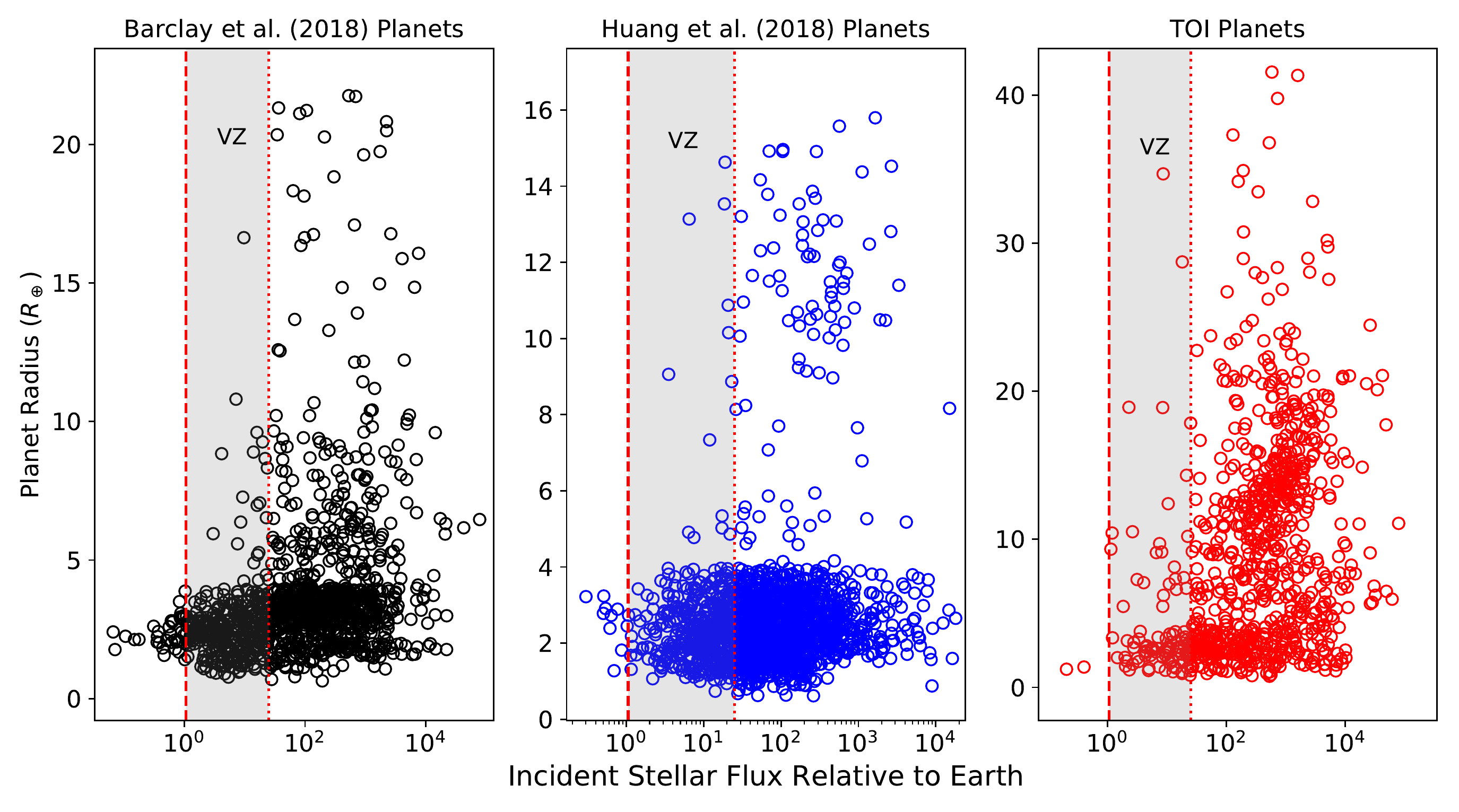}
  \caption{Comparison of the incident flux vs radius distributions of planets from the Barclay yield, Huang yield, and TOI list, demonstrating discrepancies in the predicted yield for certain planet size regimes.}
  \label{BCTComparison}
\end{figure*}

The results discussed in Section~\ref{venus yield} show that the TOI list for the first 13 sectors of the {\it TESS} primary mission contain 46 Venus analogs, which is significantly less than the 259 predicted by the Huang yield. A possible reason for this is a relatively conservative approach adopted by the {\it TESS} pipeline and vetting procedure designed to minimize false positives in the TOI yield. Therefore, we can expect an increase in the amount of planets discovered by {\it TESS} after there has been sufficient time for the community to conduct further analysis of {\it TESS} data and for the vetting process to be optimized. This is especially relevant for the Huang yield since it predicts that {\it TESS} will be finding a multitude of planets around faint stars with {\it TESS} magnitudes in the range 12--15. Furthermore, the Huang yield has so far overestimated the amount of planets with radii less than $4 \, R_\oplus$. Figure~\ref{BCTComparison} illustrates this discrepancy as it can be seen that the Huang yield underestimates the number of planets with radii larger than $4 \, R_\oplus$ that have been found in sectors 1--13.

\section{Discussion}
\label{discussion}

In this work we have conducted an analysis of both a simulated planet population generated by \citet{Huang} and the Cycle~1 TOI sample provided by the {\it TESS} mission.
It should be noted that we did not include any uncertainties related to TOI parameters, which in some cases may provide a deciding factor regarding the disposition of the planet candidate as terrestrial or gas giant. However the most significant effect is seen in the S/N values calculated using Equation~\ref{tsm equation}, as the propagation of error through this equation results in large uncertainties in the estimated S/N of TOI planets. We have also made several minor assumptions in the temperature-mass relationship used to plot the VZ, including requiring the stars used to develop the relationship to be 4 Gyrs of age, and to have a [Fe/H] metallicity between -0.1142 and 0.0142 dex (Section~\ref{Venus analog defined}). These constraints have the potential to cause the VZ boundaries shown in Figure~\ref{VZ1} to change slightly. However, our estimate on the total number of Venus analogs in the Huang yield and TOI list were made without the use of these relationships. The stellar mass and temperature relationship was only needed to plot the VZ in reference to stellar temperature, and the T-band and J-band magnitude relationship was only used to calculate TSM values.

The effects these uncertainties have on the labeling of planets as terrestrial or gaseous help promote the need for more precise measurements of stellar radii, especially for fainter stars. Our analysis of a planet is based on the extent to which we can constrain the properties of the host star. When it comes to observing a planet via the transit method, uncertainties in the stellar radius are exaggerated when attempting to constrain the radius of the planet. This will ultimately affect the target selection process, as we may avoid follow-up observations for many planets we assume to be gaseous but are in fact terrestrial.

\section{Conclusions}
\label{conclusions}

In this paper we have presented new results on the expected frequency and follow-up prospects of potential Venus analogs discovered by {\it TESS}. Our analysis of predicted {\it TESS} yields shows that the {\it TESS} mission is expected to produce $\sim$300 Venus analogs, while the TOI list from observations of sectors 1--13 contains 46 Venus analogs. We applied Equation~\ref{tsm equation} to these potential Venus analogs and found that several of the Huang yield and TOI exo-Venuses have S/N greater than 20, making it likely that absorption features would be observed in their atmospheres (Figure~\ref{TSM}). Finally, we used the Huang yield to update the S/N cutoff values developed by \citet{Kempton}, which are to be used to prioritize TESS planets for follow-up transmission spectroscopy observations (Table~\ref{Table 1}).     

The study of exoplanet analogs of Venus necessitates a collaboration of exoplanet transmission spectroscopy observations with Venusian science \citep{Kane2014venus}. Refined measurements of the atmosphere of Venus, from the upper layers down to the surface, are essential as the transparency of the upper atmosphere is the only observable portion of the atmosphere from which conditions within the deeper atmosphere and at the surface may be inferred \citep{Ehrenreich2012}. It will be difficult to detect an unambiguous Venus analog \citep{Barstow2016venus}, however observations of an inflated atmosphere could hint to a planet transitioning into a runaway greenhouse state \citep{Turbet}, and trace atmospheric constituents can be used to infer its climate using a 3-D GCM. ROCKE-3D \citep{Way_2017} is a GCM which has proven to be capable of depicting a Venus-like climate \citep{Way2018,Kane2018}, and will be a crucial tool in our attempt to characterize exo-Venuses \citep{Wolf2019}.  These models will ultimately create a diverse set of exo-climates, which will give statistical insight into the likelihood of a terrestrial planet becoming more Earth- or Venus-like, as well as refining the location of the runaway greenhouse boundary, and ultimately the VZ and HZ as a whole.


\section*{Acknowledgements}

The authors would like to thank Alma Ceja, Paul Dalba, Michelle Hill, Eliza Kempton, Zhexing Li, and George Ricker for useful feedback on the manuscript. Thanks are also due to the anonymous referee, whose comments greatly improved the quality of the paper. This research has made use of the following archives: the Habitable Zone Gallery at hzgallery.org and the NASA Exoplanet Archive, which is operated by the California Institute of Technology, under contract with the National Aeronautics and Space Administration under the Exoplanet Exploration Program. This paper includes data collected by the {\it TESS} mission, which are publicly available from the Mikulski Archive for Space Telescopes (MAST). We acknowledge the use of public {\it TESS} Alert data from pipelines at the {\it TESS} Science Office and at the {\it TESS} Science Processing Operations Center. Funding for the {\it TESS} mission is provided by NASA's Science Mission directorate. The results reported herein benefited from collaborations and/or information exchange within NASA's Nexus for Exoplanet System Science (NExSS) research coordination network sponsored by NASA's Science Mission Directorate.



\begin{thebibliography}{}
\expandafter\ifx\csname natexlab\endcsname\relax\def\natexlab#1{#1}\fi

\bibitem[{{Airapetian} {et~al.}(2017){Airapetian}, {Glocer}, {Khazanov},
  {Loyd}, {France}, {Sojka}, {Danchi}, \& {Liemohn}}]{Airapetian2019}
{Airapetian}, V.~S., {Glocer}, A., {Khazanov}, G.~V., {et~al.} 2017, \apj, 836,
  L3

\bibitem[{{Andrae} {et~al.}(2018){Andrae}, {Fouesneau}, {Creevey}, {Ordenovic},
  {Mary}, {Burlacu}, {Chaoul}, {Jean-Antoine-Piccolo}, {Kordopatis}, {Korn},
  {Lebreton}, {Panem}, {Pichon}, {Th{\'e}venin}, {Walmsley}, \&
  {Bailer-Jones}}]{Andrae2018}
{Andrae}, R., {Fouesneau}, M., {Creevey}, O., {et~al.} 2018, \aap, 616, A8

\bibitem[{{Ballard}(2019)}]{Ballard2019}
{Ballard}, S. 2019, \aj, 157, 113

\bibitem[{{Ballard} \& {Johnson}(2016)}]{Ballard2016}
{Ballard}, S., \& {Johnson}, J.~A. 2016, \apj, 816, 66

\bibitem[{{Barclay} {et~al.}(2018){Barclay}, {Pepper}, \&
  {Quintana}}]{Barclay2018}
{Barclay}, T., {Pepper}, J., \& {Quintana}, E.~V. 2018, The Astrophysical
  Journal Supplement Series, 239, 2

\bibitem[{{Barstow} {et~al.}(2015){Barstow}, {Aigrain}, {Irwin}, {Kendrew}, \&
  {Fletcher}}]{Barstow2015}
{Barstow}, J.~K., {Aigrain}, S., {Irwin}, P.~G.~J., {Kendrew}, S., \&
  {Fletcher}, L.~N. 2015, \mnras, 448, 2546

\bibitem[{{Barstow} {et~al.}(2016{\natexlab{a}}){Barstow}, {Aigrain}, {Irwin},
  {Kendrew}, \& {Fletcher}}]{Barstow2016venus}
---. 2016{\natexlab{a}}, \mnras, 458, 2657

\bibitem[{{Barstow} {et~al.}(2016{\natexlab{b}}){Barstow}, {Irwin}, {Kendrew},
  \& {Aigrain}}]{Barstow2016}
{Barstow}, J.~K., {Irwin}, P. G.~J., {Kendrew}, S., \& {Aigrain}, S.
  2016{\natexlab{b}}, in Society of Photo-Optical Instrumentation Engineers
  (SPIE) Conference Series, Vol. 9904, Space Telescopes and Instrumentation
  2016: Optical, Infrared, and Millimeter Wave, 99043P

\bibitem[{{Batalha} \& {Line}(2017)}]{BatalhaLine}
{Batalha}, N.~E., \& {Line}, M.~R. 2017, \aj, 153, 151

\bibitem[{{Beichman} {et~al.}(2014){Beichman}, {Benneke}, {Knutson}, {Smith},
  {Dressing}, {Latham}, {Deming}, {Lunine}, {Lagage}, {Sozzetti}, {Beichman},
  {Sing}, {Kempton}, {Ricker}, {Bean}, {Kreidberg}, {Bouwman}, {Crossfield},
  {Christiansen}, {Ciardi}, {Fortney}, {Albert}, {Doyon}, {Rieke}, {Rieke},
  {Clampin}, {Greenhouse}, {Goudfrooij}, {Hines}, {Keyes}, {Lee}, {McCullough},
  {Robberto}, {Stansberry}, {Valenti}, {Deroo}, {Mand ell}, {Ressler},
  {Shporer}, {Swain}, {Vasisht}, {Carey}, {Krick}, {Birkmann}, {Ferruit},
  {Giardino}, {Greene}, \& {Howell}}]{Beichman2014}
{Beichman}, C., {Benneke}, B., {Knutson}, H., {et~al.} 2014, arXiv e-prints,
  arXiv:1411.1754

\bibitem[{{Belu} {et~al.}(2011){Belu}, {Selsis}, {Morales}, {Ribas}, {Cossou},
  \& {Rauer}}]{Belu2011}
{Belu}, A.~R., {Selsis}, F., {Morales}, J.~C., {et~al.} 2011, \aap, 525, A83

\bibitem[{{Bouma} {et~al.}(2017){Bouma}, {Winn}, {Kosiarek}, \&
  {McCullough}}]{Bouma2017}
{Bouma}, L.~G., {Winn}, J.~N., {Kosiarek}, J., \& {McCullough}, P.~R. 2017,
  arXiv e-prints, arXiv:1705.08891

\bibitem[{{Boyajian} {et~al.}(2012){Boyajian}, {von Braun}, {van Belle},
  {McAlister}, {ten Brummelaar}, {Kane}, {Muirhead}, {Jones}, {White},
  {Schaefer}, {Ciardi}, {Henry}, {L{\'o}pez-Morales}, {Ridgway}, {Gies}, {Jao},
  {Rojas-Ayala}, {Parks}, {Sturmann}, {Sturmann}, {Turner}, {Farrington},
  {Goldfinger}, \& {Berger}}]{Boyajian2012}
{Boyajian}, T.~S., {von Braun}, K., {van Belle}, G., {et~al.} 2012, The
  Astrophysical Journal, 757, 112

\bibitem[{{Chen} \& {Kipping}(2017)}]{ChenKipping}
{Chen}, J., \& {Kipping}, D. 2017, \apj, 834, 17

\bibitem[{{Clampin}(2011)}]{Clampin2011}
{Clampin}, M. 2011, in IAU Symposium, Vol. 276, The Astrophysics of Planetary
  Systems: Formation, Structure, and Dynamical Evolution, ed. A.~{Sozzetti},
  M.~G. {Lattanzi}, \& A.~P. {Boss}, 335--342

\bibitem[{{Croll} {et~al.}(2011){Croll}, {Albert}, {Jayawardhana},
  {Miller-Ricci Kempton}, {Fortney}, {Murray}, \& {Neilson}}]{Croll2011}
{Croll}, B., {Albert}, L., {Jayawardhana}, R., {et~al.} 2011, \apj, 736, 78

\bibitem[{{Crouzet} {et~al.}(2017){Crouzet}, {Bonfils}, {Delfosse}, {Boisse},
  {H{\'e}brard}, {Forveille}, {Donati}, {Bouchy}, {Moutou}, {Doyon}, {Artigau},
  {Albert}, {Malo}, {Lecavelier des Etangs}, \& {Santerne}}]{Crouzet2017}
{Crouzet}, N., {Bonfils}, X., {Delfosse}, X., {et~al.} 2017, arXiv e-prints,
  arXiv:1701.03539

\bibitem[{{Deming} {et~al.}(2009){Deming}, {Seager}, {Winn}, {Miller-Ricci},
  {Clampin}, {Lindler}, {Greene}, {Charbonneau}, {Laughlin}, {Ricker},
  {Latham}, \& {Ennico}}]{Deming2009}
{Deming}, D., {Seager}, S., {Winn}, J., {et~al.} 2009, Publications of the
  Astronomical Society of the Pacific, 121, 952

\bibitem[{{Dressing} \& {Charbonneau}(2015)}]{DressingCharb}
{Dressing}, C.~D., \& {Charbonneau}, D. 2015, \apj, 807, 45

\bibitem[{{Ehrenreich} {et~al.}(2012){Ehrenreich}, {Vidal-Madjar}, {Widemann},
  {Gronoff}, {Tanga}, {Barth{\'e}lemy}, {Lilensten}, {Lecavelier Des Etangs},
  \& {Arnold}}]{Ehrenreich2012}
{Ehrenreich}, D., {Vidal-Madjar}, A., {Widemann}, T., {et~al.} 2012, \aap, 537,
  L2

\bibitem[{{Eker} {et~al.}(2015){Eker}, {Soydugan}, {Soydugan}, {Bilir}, {Yaz
  G{\"o}k{\c{c}}e}, {Steer}, {T{\"u}ys{\"u}z}, {{\c{S}}eny{\"u}z}, \&
  {Demircan}}]{Eker2015}
{Eker}, Z., {Soydugan}, F., {Soydugan}, E., {et~al.} 2015, The Astronomical
  Journal, 149, 131

\bibitem[{{Fressin} {et~al.}(2013){Fressin}, {Torres}, {Charbonneau}, {Bryson},
  {Christiansen}, {Dressing}, {Jenkins}, {Walkowicz}, \&
  {Batalha}}]{Fressin2013}
{Fressin}, F., {Torres}, G., {Charbonneau}, D., {et~al.} 2013, \apj, 766, 81

\bibitem[{{Gillon} {et~al.}(2017){Gillon}, {Triaud}, {Demory}, {Jehin}, {Agol},
  {Deck}, {Lederer}, {de Wit}, {Burdanov}, {Ingalls}, {Bolmont}, {Leconte},
  {Raymond}, {Selsis}, {Turbet}, {Barkaoui}, {Burgasser}, {Burleigh}, {Carey},
  {Chaushev}, {Copperwheat}, {Delrez}, {Fernand es}, {Holdsworth}, {Kotze},
  {Van Grootel}, {Almleaky}, {Benkhaldoun}, {Magain}, \& {Queloz}}]{Gillon2017}
{Gillon}, M., {Triaud}, A. H.~M.~J., {Demory}, B.-O., {et~al.} 2017, \nat, 542,
  456

\bibitem[{{Greene} {et~al.}(2016){Greene}, {Line}, {Montero}, {Fortney},
  {Lustig-Yaeger}, \& {Luther}}]{Greene2016}
{Greene}, T.~P., {Line}, M.~R., {Montero}, C., {et~al.} 2016, \apj, 817, 17

\bibitem[{{Howe} {et~al.}(2017){Howe}, {Burrows}, \& {Deming}}]{Howe2017}
{Howe}, A.~R., {Burrows}, A., \& {Deming}, D. 2017, \apj, 835, 96

\bibitem[{{Huang} {et~al.}(2018){Huang}, {Shporer}, {Dragomir}, {Fausnaugh},
  {Levine}, {Morgan}, {Nguyen}, {Ricker}, {Wall}, {Woods}, \&
  {Vanderspek}}]{Huang}
{Huang}, C.~X., {Shporer}, A., {Dragomir}, D., {et~al.} 2018, arXiv e-prints,
  arXiv:1807.11129

\bibitem[{{Kane} {et~al.}(2018){Kane}, {Ceja}, {Way}, \& {Quintana}}]{Kane2018}
{Kane}, S.~R., {Ceja}, A.~Y., {Way}, M.~J., \& {Quintana}, E.~V. 2018, \apj,
  869, 46

\bibitem[{{Kane} {et~al.}(2014){Kane}, {Kopparapu}, \&
  {Domagal-Goldman}}]{Kane2014venus}
{Kane}, S.~R., {Kopparapu}, R.~K., \& {Domagal-Goldman}, S.~D. 2014, \apj, 794,
  L5

\bibitem[{{Kane} {et~al.}(2009){Kane}, {Mahadevan}, {von Braun}, {Laughlin}, \&
  {Ciardi}}]{Kane2009}
{Kane}, S.~R., {Mahadevan}, S., {von Braun}, K., {Laughlin}, G., \& {Ciardi},
  D.~R. 2009, Publications of the Astronomical Society of the Pacific, 121,
  1386

\bibitem[{{Kane} {et~al.}(2019){Kane}, {Arney}, {Crisp}, {Domagal-Goldman2},
  {Glaze}, {Goldblatt}, {Grinspoon}, {Head}, {Lenardic}, {Unterborn}, {Way}, \&
  {Zahnle}}]{Kane2019}
{Kane}, S.~R., {Arney}, G., {Crisp}, D., {et~al.} 2019, arXiv e-prints,
  arXiv:1908.02783

\bibitem[{{Kempton} {et~al.}(2018){Kempton}, {Bean}, {Louie}, {Deming}, {Koll},
  {Mansfield}, {Christiansen}, {L{\'o}pez-Morales}, {Swain}, {Zellem},
  {Ballard}, {Barclay}, {Barstow}, {Batalha}, {Beatty}, {Berta-Thompson},
  {Birkby}, {Buchhave}, {Charbonneau}, {Cowan}, {Crossfield}, {de Val-Borro},
  {Doyon}, {Dragomir}, {Gaidos}, {Heng}, {Hu}, {Kane}, {Kreidberg}, {Mallonn},
  {Morley}, {Narita}, {Nascimbeni}, {Pall{\'e}}, {Quintana}, {Rauscher},
  {Seager}, {Shkolnik}, {Sing}, {Sozzetti}, {Stassun}, {Valenti}, \& {von
  Essen}}]{Kempton}
{Kempton}, E. M.~R., {Bean}, J.~L., {Louie}, D.~R., {et~al.} 2018, Publications
  of the Astronomical Society of the Pacific, 130, 114401

\bibitem[{{Kislyakova} {et~al.}(2019){Kislyakova}, {Holmstr{\"o}m}, {Odert},
  {Lammer}, {Erkaev}, {Khodachenko}, {Shaikhislamov}, {Dorfi}, \&
  {G{\"u}del}}]{Kislyakova2019}
{Kislyakova}, K.~G., {Holmstr{\"o}m}, M., {Odert}, P., {et~al.} 2019, \aap,
  623, A131

\bibitem[{{Kopparapu} {et~al.}(2013){Kopparapu}, {Ramirez}, {Kasting}, {Eymet},
  {Robinson}, {Mahadevan}, {Terrien}, {Domagal-Goldman}, {Meadows}, \&
  {Deshpande}}]{Kopparapu2013}
{Kopparapu}, R.~K., {Ramirez}, R., {Kasting}, J.~F., {et~al.} 2013, \apj, 765,
  131

\bibitem[{{Kreidberg} {et~al.}(2016){Kreidberg}, {Morley}, {Line}, {Stevenson},
  \& {Dragomir}}]{Kreidberg2016}
{Kreidberg}, L., {Morley}, C., {Line}, M., {Stevenson}, K., \& {Dragomir}, D.
  2016, {Clouds in the Forecast? A Joint Spitzer and HST Investigation of
  Clouds and Hazes for Two Exo-Neptunes}, Spitzer Proposal, ,

\bibitem[{{Lingam} \& {Loeb}(2018)}]{LingLoeb2017a}
{Lingam}, M., \& {Loeb}, A. 2018, International Journal of Astrobiology, 17,
  116

\bibitem[{{Lopez} \& {Fortney}(2014)}]{Lopez2014}
{Lopez}, E.~D., \& {Fortney}, J.~J. 2014, \apj, 792, 1

\bibitem[{{Louie} {et~al.}(2018){Louie}, {Deming}, {Albert}, {Bouma}, {Bean},
  \& {Lopez-Morales}}]{Louie2018}
{Louie}, D.~R., {Deming}, D., {Albert}, L., {et~al.} 2018, Publications of the
  Astronomical Society of the Pacific, 130, 044401

\bibitem[{{Molli{\`e}re} {et~al.}(2017){Molli{\`e}re}, {van Boekel}, {Bouwman},
  {Henning}, {Lagage}, \& {Min}}]{Molliere2017}
{Molli{\`e}re}, P., {van Boekel}, R., {Bouwman}, J., {et~al.} 2017, \aap, 600,
  A10

\bibitem[{{Ricker} {et~al.}(2015){Ricker}, {Winn}, {Vanderspek}, {Latham},
  {Bakos}, {Bean}, {Berta-Thompson}, {Brown}, {Buchhave}, {Butler}, {Butler},
  {Chaplin}, {Charbonneau}, {Christensen-Dalsgaard}, {Clampin}, {Deming},
  {Doty}, {De Lee}, {Dressing}, {Dunham}, {Endl}, {Fressin}, {Ge}, {Henning},
  {Holman}, {Howard}, {Ida}, {Jenkins}, {Jernigan}, {Johnson}, {Kaltenegger},
  {Kawai}, {Kjeldsen}, {Laughlin}, {Levine}, {Lin}, {Lissauer}, {MacQueen},
  {Marcy}, {McCullough}, {Morton}, {Narita}, {Paegert}, {Palle}, {Pepe},
  {Pepper}, {Quirrenbach}, {Rinehart}, {Sasselov}, {Sato}, {Seager},
  {Sozzetti}, {Stassun}, {Sullivan}, {Szentgyorgyi}, {Torres}, {Udry}, \&
  {Villasenor}}]{Ricker2015}
{Ricker}, G.~R., {Winn}, J.~N., {Vanderspek}, R., {et~al.} 2015, Journal of
  Astronomical Telescopes, Instruments, and Systems, 1, 014003

\bibitem[{{Rogers}(2015)}]{Rogers2015}
{Rogers}, L.~A. 2015, \apj, 801, 41

\bibitem[{Smrekar {et~al.}(2010)Smrekar, Stofan, Mueller, Treiman,
  Elkins-Tanton, Helbert, Piccioni, \& Drossart}]{Smrekar605}
Smrekar, S.~E., Stofan, E.~R., Mueller, N., {et~al.} 2010, Science, 328, 605

\bibitem[{{Stassun} {et~al.}(2018){Stassun}, {Oelkers}, {Pepper}, {Paegert},
  {De Lee}, {Torres}, {Latham}, {Charpinet}, {Dressing}, {Huber}, {Kane},
  {L{\'e}pine}, {Mann}, {Muirhead}, {Rojas-Ayala}, {Silvotti}, {Fleming},
  {Levine}, \& {Plavchan}}]{Stassun2018}
{Stassun}, K.~G., {Oelkers}, R.~J., {Pepper}, J., {et~al.} 2018, \aj, 156, 102

\bibitem[{Stevenson {et~al.}(2016)Stevenson, Lewis, Bean, Beichman, Fraine,
  Kilpatrick, Krick, Lothringer, Mandell, Valenti,
  {et~al.}}]{stevenson2016transiting}
Stevenson, K.~B., Lewis, N.~K., Bean, J.~L., {et~al.} 2016, Publications of the
  Astronomical Society of the Pacific, 128, 094401

\bibitem[{{Sullivan} {et~al.}(2015){Sullivan}, {Winn}, {Berta-Thompson},
  {Charbonneau}, {Deming}, {Dressing}, {Latham}, {Levine}, {McCullough},
  {Morton}, {Ricker}, {Vanderspek}, \& {Woods}}]{Sullivan2015}
{Sullivan}, P.~W., {Winn}, J.~N., {Berta-Thompson}, Z.~K., {et~al.} 2015, \apj,
  809, 77

\bibitem[{{Turbet} {et~al.}(2019){Turbet}, {Ehrenreich}, {Lovis}, {Bolmont}, \&
  {Fauchez}}]{Turbet}
{Turbet}, M., {Ehrenreich}, D., {Lovis}, C., {Bolmont}, E., \& {Fauchez}, T.
  2019, arXiv e-prints, arXiv:1906.03527

\bibitem[{{van Saders} {et~al.}(2016){van Saders}, {Ceillier}, {Metcalfe},
  {Silva Aguirre}, {Pinsonneault}, {Garc{\'\i}a}, {Mathur}, \&
  {Davies}}]{vansaders2016}
{van Saders}, J.~L., {Ceillier}, T., {Metcalfe}, T.~S., {et~al.} 2016, \nat,
  529, 181

\bibitem[{{Way} {et~al.}(2018){Way}, {Del Genio}, \& {Amundsen}}]{Way2018}
{Way}, M.~J., {Del Genio}, A., \& {Amundsen}, D.~S. 2018, arXiv e-prints,
  arXiv:1802.05434

\bibitem[{{Way} {et~al.}(2016){Way}, {Del Genio}, {Kiang}, {Sohl}, {Grinspoon},
  {Aleinov}, {Kelley}, \& {Clune}}]{Way2016}
{Way}, M.~J., {Del Genio}, A.~D., {Kiang}, N.~Y., {et~al.} 2016, \grl, 43, 8376

\bibitem[{Way {et~al.}(2017)Way, Aleinov, Amundsen, Chandler, Clune, Genio,
  Fujii, Kelley, Kiang, Sohl, \& Tsigaridis}]{Way_2017}
Way, M.~J., Aleinov, I., Amundsen, D.~S., {et~al.} 2017, The Astrophysical
  Journal Supplement Series, 231, 12

\bibitem[{{Wolf} {et~al.}(2019){Wolf}, {Kopparapu}, {Airapetian}, {Fauchez},
  {Guzewich}, {Kane}, {Pidhorodetska}, {Way}, {Abbot}, {Checlair}, {Davis},
  {Del Genio}, {Dong}, {Eggl}, {Fleming}, {Fujii}, {Haghighipour}, {Heavens},
  {Henning}, {Kiang}, {Lopez-Morales}, {Lustig-Yaeger}, {Meadows}, {Reinhard},
  {Rugheimer}, {Schwieterman}, {Shields}, {Sohl}, {Turbet}, \&
  {Wordsworth}}]{Wolf2019}
{Wolf}, E.~T., {Kopparapu}, R., {Airapetian}, V., {et~al.} 2019, arXiv
  e-prints, arXiv:1903.05012

\bibitem[{{Zahnle} \& {Catling}(2017)}]{ZahnleCatling}
{Zahnle}, K.~J., \& {Catling}, D.~C. 2017, \apj, 843, 122

\bibitem[{{Zendejas} {et~al.}(2010){Zendejas}, {Segura}, \&
  {Raga}}]{Zendejas2010}
{Zendejas}, J., {Segura}, A., \& {Raga}, A.~C. 2010, \icarus, 210, 539

\bibitem[{{Zhu} {et~al.}(2018){Zhu}, {Petrovich}, {Wu}, {Dong}, \&
  {Xie}}]{Zhu2018}
{Zhu}, W., {Petrovich}, C., {Wu}, Y., {Dong}, S., \& {Xie}, J. 2018, \apj, 860,
  101

\end{thebibliography}


\end{document}